\documentclass[twocolumn,showpacs,preprintnumbers,amsmath,amssymb]{revtex4}
\usepackage{graphicx}% Include figure files
\usepackage{dcolumn}% Align table columns on decimal point
\usepackage{bm}% bold math

\begin{document}

%\preprint{CPTh/PC 006.0106}

\title{Infrared finite coupling in Sudakov resummation: the precise set-up}% Force line breaks with \\

\author{Georges Grunberg}
%\email{grunberg@cpht.polytechnique.fr}
\affiliation{%
Centre de Physique Th\'eorique, Ecole  
Polytechnique, CNRS\\
        91128 Palaiseau Cedex, France
}%

%\affiliation{

\date{\today}% It is always \today, today,

\begin{abstract}
I show that Sudakov resummation takes a  transparent form if one deals with
the second logarithmic derivative of the short distance coefficient functions for deep inelastic scattering and 
the Drell-Yan process. A uniquely defined Sudakov exponent emerges, and  the  constant
terms not included in the exponent are conjectured to be  given by the second logarithmic derivative of the
massless quark form factor. The precise framework for the implementation of the dispersive approach
to power corrections is set-up, yielding results in agreement with  infrared renormalon expectations, but which
are not tied to the  single (dressed) gluon exchange approximation.  Indications for a Banks-Zaks type of 
perturbative fixed point in the Sudakov effective coupling at low $N_f$ are  pointed out. Existence of a fixed
point in the Sudakov coupling  is argued to imply its  universality.  
\end{abstract}

\maketitle

\noindent The infrared (IR) finite coupling (``dispersive'') approach to power corrections \cite{DMW} provides an
attractive framework where the issue of universality can be meaningfully raised. This approach however seems to be
tied in an essential way to the single gluon exchange approximation. In this paper (which is an improved version of
\cite{Gru-talk2}) I show that it can actually find a precise implementation in the framework of Sudakov
resummation, and that its validity extends beyond single gluon exchange. Previously discussed
\cite{Gru-talk,Gru-talk1} ambiguities, related to the variety of resummation procedures, are now resolved through
an additional prescription. 
Some new results are presented concerning the universality issue. It also appears that a previously provided 
\cite{Gru-talk,Gru-talk1} large $N_f$ evidence in favor of the IR finite coupling approach should be dismissed, the
coupling having been incorrectly identified as Euclidean. The new interpretation of the same formal results still
supports the IR finite coupling idea, but in a non-perturbative framework, in the spirit of the original
\cite{DMW} proposal. Some indications for an IR finite  perturbative Euclidean
coupling of the Banks-Zaks type at rather {\em small} $N_f$   are nevertheless pointed out.

Consider first the scaling violation in deep inelastic scattering (DIS) in Mellin space at large $N$. One can show
\cite{Gru-talk,Gru-talk1} that Sudakov resummation takes in this case the very simple form

\begin{eqnarray}{d\ln F_2(Q^2,N)\over d\ln Q^2}=4 C_F\int_{0}^{Q^2}{dk^2\over k^2} G(N k^2/ Q^2)
A_{{\cal S}}(k^2)\nonumber\\
+4 C_F H(Q^2)+{\cal O}(1/ N)  
\label{eq:scale-viol},\end{eqnarray}
where   the ``Sudakov effective coupling'' $A_{{\cal S}}(k^2)=a_s(k^2)+{\cal
A}_1 a_s^2(k^2)+{\cal A}_2 a_s^3(k^2)+...$, as well as $H(Q^2)=h_0 a_s(Q^2)+h_1 a_s^2(Q^2)+h_2
a_s^3(Q^2)+...$,  are  given as  power series in
$a_s\equiv \alpha_s/4\pi$ with $N$-independent coefficients. In the standard resummation framework one has
$A_{{\cal S}}(k^2)=A_{{\cal S}}^{stan}(k^2)$ with

\begin{equation}4 C_F A_{{\cal S}}^{stan}(k^2)=A(a_s(k^2))+dB(a_s(k^2))/d\ln
k^2
\label{eq:A-S},\end{equation}
 where
$A$ (the universal  ``cusp'' anomalous dimension) and $B$ are  the standard Sudakov anomalous dimensions
relevant to DIS, and
$G(N k^2/Q^2)=G_{stan}(N k^2/Q^2)\equiv \exp(-N k^2/Q^2)-1$.
The Sudakov integral on the right hand side of eq.(\ref{eq:scale-viol}) contains, besides logarithmic and constant
terms, also
 terms which vanish at large $N$ order by order in perturbation theory. The latter can be removed, and
absorbed into the  ${\cal O}(1/N)$ terms, using the equivalent relation 

\begin{eqnarray}{d\ln F_2(Q^2,N)\over d\ln Q^2}=4 C_F\Big[\int_{0}^{\infty}{dk^2\over k^2} G(N k^2/ Q^2)
A_{{\cal S}}(k^2)\nonumber\\
- G(\infty)\int_{Q^2}^{\infty}{dk^2\over k^2} A_{{\cal
S}}(k^2)\Big]\nonumber\\
+4 C_F H(Q^2)+{\cal O}(1/ N)\label{eq:ren-int-scaling-as}.\end{eqnarray}
The two integrals on the right hand side of eq.(\ref{eq:ren-int-scaling-as}), when expanded in powers of
$a_s(Q^2)$, contain only
logarithmic and constant terms, and  are free of ${\cal O}(1/N)$ terms. Since
$G(\infty)=-1$, these integrals  are separately ultraviolet (UV)
divergent, but their sum is finite. It was further  observed in \cite{Gru-talk,Gru-talk1} that the separation
between the constant terms contained in the Sudakov integrals on the right hand side of eq.(\ref{eq:scale-viol})
 or (\ref{eq:ren-int-scaling-as}) and the ``leftover'' constant terms contained in $H(Q^2)$ is arbitrary,
yielding a variety of Sudakov resummation  procedures, different choices leading to a different ``Sudakov
distribution function''
$G(N k^2/Q^2)$ and effective coupling
$A_{{\cal S}}(k^2)$, as well as to a different function
$H(Q^2)$.   The  new 
observation of the present paper is  that this  freedom of selecting the constant terms actually  disappears by
taking one more derivative, namely

\begin{eqnarray}{d^2\ln F_2(Q^2,N)\over (d\ln Q^2)^2}=4 C_F\int_{0}^{\infty}{dk^2\over k^2}
\dot{G}(N k^2/ Q^2) A_{{\cal S}}(k^2)\nonumber\\
+4 C_F[dH/ d\ln Q^2-A_{{\cal S}}(Q^2)]+{\cal
O}(1/ N)\label{eq:d-scale-viol},\end{eqnarray}
where $\dot{G}=- dG/ d\ln k^2$. The point is that the integral on the right hand side of
eq.(\ref{eq:d-scale-viol}) being UV convergent, all the large $N$ logarithmic terms are now determined by the
${\cal O}(N^0)$ terms contained in the integral, which therefore cannot be fixed arbitrarily anymore.
Indeed, putting 
\begin{equation}{\cal S}(Q^2/N)=\int_{0}^{\infty}{dk^2\over k^2}
\dot{G}(N k^2/ Q^2) A_{{\cal S}}(k^2)\label{eq:S-exp},\end{equation}
it is easy to show that

\begin{eqnarray}{\cal S}(Q^2/N)&=c_0 a_s(Q^2)+(\beta_0 c_0 L + {\cal A}_1 c_0-\beta_0 c_1)
a_s^2(Q^2)\nonumber\\
&+[\beta_0^2 c_0 L^2 +(2\beta_0 ({\cal A}_1 c_0-\beta_0 c_1)+\beta_1 c_0)L\nonumber\\
&+{\cal O}(L^0)]a_s^3(Q^2)
+...
\label{eq:log-structure},\end{eqnarray}
where $L=\ln N$ and $c_p=\int_{0}^{\infty}{d\epsilon\over \epsilon} \dot{G}(\epsilon)\ln^p(\epsilon)$. Thus $c_0$
determines all the leading logarithms of $N$ (which implies that $c_0=1$), while the combination  ${\cal A}_1
c_0-\beta_0 c_1$ determines the sub-leading logarithms and is therefore fixed, etc..., which shows that the
Sudakov exponent ${\cal S}(Q^2/N)$ is  uniquely determined.
 
\noindent This observation implies in turn that  the combination
$dH/ d\ln Q^2-A_{{\cal S}}(Q^2)$, which represents the ``leftover'' constant terms not included in ${\cal
S}(Q^2/N)$, is also uniquely fixed. In fact, I conjecture that it is related to the space-like on-shell
electromagnetic quark form factor
\cite{Magnea-Sterman,MVV} ${\cal F}_q(Q^2)$ by
  
\begin{equation}4 C_F\left({dH\over d\ln Q^2}-A_{{\cal S}}(Q^2)\right)={d^2\ln \left({\cal F}_q(Q^2)\right)^2 \over
(d\ln Q^2)^2}\label{eq:form-factor}.\end{equation}
The fact that the second logarithmic derivative of the form factor is both finite and renormalization group
invariant follows from the properties \cite{Collins} of the evolution equation   satisfied by the form factor.
Eq.(\ref{eq:form-factor}) has been checked \cite{Gru-Friot} to  ${\cal O}(a_s^4)$. Further checks  to all orders
at large $N_f$ are  under consideration. Eq.(\ref{eq:form-factor}) shows  that the choice of $A_{{\cal S}}$ is
indeed correlated with that of $H$, whose derivative represents the ``non-universal'' part of $A_{{\cal S}}$.
Thus Sudakov resummation takes the very suggestive form

\begin{eqnarray}{d^2\ln F_2(Q^2,N)\over (d\ln Q^2)^2}=4 C_F{\cal S}(Q^2/N)+{d^2\ln \left({\cal F}_q(Q^2)\right)^2 
\over(d\ln Q^2)^2}\nonumber\\
+{\cal O}(1/N)\label{eq:S+form-factor},\end{eqnarray}
where the form factor term  isolates the virtual contributions.

\noindent For the short distance Drell-Yan (DY) cross
section, the analogues of eq.(\ref{eq:scale-viol}),  (\ref{eq:d-scale-viol}) and (\ref{eq:form-factor}) are

\begin{eqnarray}{d\ln \sigma_{DY}(Q^2,N)\over d\ln Q^2}&=4 C_F\int_{0}^{Q^2}{dk^2\over k^2} G_{DY}(N k/ Q)
A_{{\cal S},DY}(k^2)\nonumber\\
&+4 C_F H_{DY}(Q^2)
+{\cal O}(1/N)  
\label{eq:scale-viol-DY},\end{eqnarray}

\begin{eqnarray}{d^2\ln \sigma_{DY}(Q^2,N)\over (d\ln Q^2)^2}&=4 C_F\int_{0}^{\infty}{dk^2\over k^2}
\dot{G}_{DY}(N k/ Q) A_{{\cal S},DY}(k^2)\nonumber\\
&+4 C_F[dH_{DY}/ d\ln Q^2-A_{{\cal S},DY}(Q^2)]\nonumber\\
&+{\cal
O}(1/ N)\label{eq:d-scale-viol-DY},\end{eqnarray}
where $\dot{G}_{DY}=- dG_{DY}/ d\ln k^2$,  
and

\begin{equation}4 C_F\left({dH_{DY}\over d\ln Q^2}-A_{{\cal S},DY}(Q^2)\right)={d^2\ln \vert{\cal F}_q(-Q^2)\vert^2
\over (d\ln Q^2)^2}\label{eq:form-factor-DY},\end{equation}
where ${\cal F}_q(-Q^2)$ is the time-like quark form factor. Within the standard resummation framework
  $G_{DY}(N k/Q)=G_{DY}^{stan}(N k/Q)\equiv\exp(-N
k/Q)-1$ and $A_{{\cal S},DY}(k^2)=A_{{\cal S},DY}^{stan}(k^2)$ with

\begin{equation}4 C_F A_{{\cal S},DY}^{stan}(k^2)=  A(a_s(k^2))+{1\over 2}{dD(a_s(k^2))\over d\ln
k^2}\label{eq:A-S-DY},\end{equation}
where $D$ is the usual D-term.
Eq.(\ref{eq:form-factor}) and
(\ref{eq:form-factor-DY}) are akin to  previous results \cite{Sterman, Magnea, MV, L-M, Ravindran} relating
$N$-independent terms
 to form factor type contributions.

However, although the
Sudakov exponent
${\cal S}(Q^2/N)$ (eq.(\ref{eq:S-exp})) is  uniquely determined, the Sudakov distribution function and effective
coupling are still {\em not}. For instance, eq.(\ref{eq:log-structure}) shows that the value of ${\cal A}_1$ is
correlated to that of
$c_1$, with only the combination ${\cal A}_1
c_0-\beta_0 c_1$ fixed. In fact, one can look at eq.(\ref{eq:S-exp}) for any given choice of the
 distribution function
$G(N k^2/Q^2)$  as defining an  integral transform mapping the  effective coupling $A_{{\cal
S}}(k^2)$ to the Sudakov exponent \cite{footnote0} ${\cal S}(Q^2/N)$. The only obvious constraint on the transform
is the normalization
$c_0=1$. From this point of view, all choices of $G$ are equivalent for
the purpose of resumming Sudakov logarithms, and the very question \cite{Gru-talk1}
whether
$A_{{\cal S}}(k^2)$ should be identified to an Euclidean or to a Minkowskian coupling cannot be answered at this
stage: we simply deal with an infinite variety of different representations of ${\cal S}(Q^2/N)$.
A prescription to  single out the correct representation  relevant for the issue of power corrections in the IR
finite coupling approach is needed, which can only be provided by additional physical information.   It is true that in principle the different mathematical  representations of 
${\cal S}(Q^2/N)$ are all equivalent. However, some nonperturbative features of $A_{{\cal S}}(k^2)$, which should
be taken into account in phenomenological parametrizations, will  depend upon  the particular representation
chosen. For instance, some choices may be incompatible with the IR finite coupling hypothesis, or may involve
renormalons in the perturbative series for
$A_{{\cal S}}(k^2)$ itself, or   introduce  large UV power corrections in
$A_{{\cal S}}(k^2)$. In practice \cite{DMW}, one would like to deal with an IR finite and renormalon-free {\em
Euclidean}
$A_{{\cal S}}(k^2)$ with no large UV power corrections. Only then the  power corrections are correctly
obtained from the low $k^2$ expansion of the Sudakov distribution function. In absence of
the appropriate criterion,  predictions such as existence of an ${\cal O}(1/Q)$ linear power correction
\cite{K-S} in Drell-Yan, or logarithmically-enhanced power corrections \cite{Gru-talk1}, which follow from 
particular choices of
$G$, cannot be a priori dismissed.
 One should also   stress that the
ansatz for the IR finite behavior of 
$A_{{\cal S}}(k^2)$ at low scales represents an alternative to the shape
function \cite{Korchemsky} approach. Moreover, 
 there is  no problem \cite{footnote} with such an ansatz to invert the Mellin
transform to momentum space.

\noindent One way to give a physical interpretation to the Sudakov effective coupling
$A_{{\cal S}}(k^2)$, and eliminate the non-uniqueness in its definition,
is  to try to
associate it to some kind of dressed gluon propagator, which does not seem possible in general,
since  resummation formulas such as eq.(\ref{eq:d-scale-viol}) or (\ref{eq:d-scale-viol-DY})  are valid beyond the
single gluon exchange approximation. However, in the large $N_f$ (``large $\beta_0$'') limit of QCD,  the
latter approximation naturally arises. In particular,  the dispersive approach \cite{DMW,BB} allows to identify a
physical {\em Minkowskian} coupling

\begin{equation}A_{{\cal
S},\infty}^{Mink}(k^2)={1\over\beta_0}\left[{1\over
2}-{1\over\pi}\arctan(t/\pi)\right]\label{eq:A-simple},\end{equation}
which is the time-like (integrated) discontinuity of the {\em Euclidean} one-loop
coupling (the  ``V-scheme'' coupling) associated to the  dressed gluon propagator

\begin{equation}A_{{\cal S},\infty}^{Eucl}(k^2)=1/ (\beta_0 t)\label{eq:one-loop},\end{equation}
where $t=\ln (k^2/\Lambda^2_V)$ ($\Lambda_V$ is the V-scheme scale parameter). It is then natural to select
the Sudakov distribution function by the requirement that the associated  effective coupling is  just
$A_{{\cal S},\infty}^{Mink}(k^2)$ at {\em large} $N_f$. As shown in \cite{Gru-talk,Gru-talk1}, this ansatz
fixes the corresponding ``Minkowskian''   distribution function (which one could also call ``characteristic
function'' following
\cite{DMW}) to be given in the DIS case by
$G_{Mink}(N k^2/Q^2)=
\ddot{{\cal G}}(\epsilon)$, with
$\epsilon=N k^2/Q^2$ and
\begin{equation} \ddot{{\cal G}}(\epsilon)=G_{stan}(\epsilon)-{1\over 2} \epsilon\
\exp(-\epsilon)-{1\over 2} \epsilon\
\Gamma(0,\epsilon)+{1\over 2} \epsilon^2\
\Gamma(0,\epsilon)
\label{eq:Gdot-SDG},\end{equation}
where $\Gamma(0,\epsilon)$ is the incomplete gamma function. The function $\ddot{{\cal G}}(\epsilon)$ is
obtained from the finite $N$ characteristic function \cite{DMW} ${\cal F}(\lambda^2/Q^2,N)$ (where
$\lambda$ is the ``gluon mass'')
 by defining
${\cal G}(y,N)\equiv{\cal F}(\lambda^2/Q^2,N)$ with $y\equiv N \lambda^2/Q^2$, and taking the
$N\rightarrow\infty$ limit at {\em fixed} $y$:
$\ddot{{\cal G}}(y,N)\rightarrow \ddot{{\cal G}}(y,\infty)\equiv
\ddot{{\cal G}}(y)$ (where $\dot{{\cal G}}=- d{\cal G}/ d\ln Q^2$).

\noindent In the Drell-Yan case, the same requirement yields 
instead $G_{Mink}^{DY}(N k/Q)=\ddot{{\cal G}}_{DY}(\epsilon_{DY}^2)$, with $\epsilon_{DY}=N k/Q$ and

\begin{equation}\ddot{{\cal G}}_{DY}(\epsilon_{DY}^2)=2 {d\over d\ln Q^2}\left[K_0(2
N k/Q)+\ln(N k/Q)+\gamma_E\right]
\label{eq:G-DY-new},\end{equation}
where $K_0$ is the modified Bessel function of the second kind, and the right hand side is an {\em even} function
of $N k/Q$. Similarly,
$\ddot{{\cal G}}_{DY}(y_{DY})$ is obtained by taking the large $N$ limit at fixed $y_{DY}= N^2 \lambda^2/Q^2$ of
the finite $N$ characteristic function
\cite{DMW} ${\cal F}_{DY}(\lambda^2/Q^2,N)$.   The same  distribution function $G_{Mink}^{DY}(N k/Q)$  also
follows from the resummation formalism (not tied to the single gluon approximation) of
\cite{Vogelsang}, which  therefore uses an implicitly Minkowskian framework in the above sense.
 
 In the Minkowskian formalism, only {\em non-analytic} terms in the small ``gluon mass'' expansion of the
characteristic function do contribute \cite{DMW} to the IR power corrections, through their discontinuities. The IR
power corrections are actually more simply parametrized
\cite{Gru-power,Euclidean}  by low energy moments of the  Euclidean coupling,  related to 
the Minkowskian coupling by the
dispersion relation

\begin{equation} A_{{\cal S}}^{Eucl}(k^2)=k^2\int_0^{\infty} dy {A_{{\cal S}}^{Mink}(y)\over (k^2+y)^2}
\label{eq:disp-coupling}.\end{equation}
It is therefore  useful
to introduce the corresponding Euclidean  distribution function $G_{Eucl}$ (where all the terms of the low
energy expansion contribute to the IR power corrections). In the DIS case, the latter should be related  to
$G_{Mink}$ by the dispersion relation

\begin{equation} G_{Mink}(\epsilon)=\epsilon\int_0^{\infty} dy {G_{Eucl}(y)\over (\epsilon+y)^2}
\label{eq:disp}.\end{equation}
Actually $G_{Eucl}$ turns out not to  exist in this case, 
since the discontinuity of the corresponding $G_{Mink}(\epsilon)$ 
(eq.(\ref{eq:Gdot-SDG})) for any finite
$\epsilon<0$  is of the form
$a \epsilon^2+b \epsilon^4$ (yielding only two \cite{Gardi-Roberts} power corrections in the Sudakov exponent,
with no logarithmic enhancement). One   then has to rely on the procedure of \cite{Gru-power,Euclidean} to relate
IR power corrections to moments of the Euclidean coupling.

\noindent In the Drell-Yan case,
one gets 
\begin{equation}G_{Eucl}^{DY}(\epsilon_{DY})=J_0(2\epsilon_{DY})-1\equiv
\tilde{G}_{Eucl}^{DY}(\epsilon_{DY}^2)\label{eq:G-DY-Eucl},\end{equation}  where $J_0$ is the Bessel function, an
{\em even} function of $\epsilon_{DY}=N k/Q$: this property ensures only even power corrections (with no
logarithmic enhancement) are present, in agreement with the renormalon argument
\cite{Beneke-Braun}. In particular, there are no ${\cal O}(1/Q)$ correction.
The analogue of eq.(\ref{eq:disp}) is

\begin{equation} \tilde{G}_{Mink}^{DY}(\epsilon_{DY}^2)=\epsilon_{DY}^2\int_0^{\infty} dy
{\tilde{G}_{Eucl}^{DY}(y)\over (\epsilon_{DY}^2+y)^2}
\label{eq:disp-DY}.\end{equation}
where  $\tilde{G}_{Mink}^{DY}(\epsilon_{DY}^2)\equiv\ddot{{\cal
G}}_{DY}(\epsilon_{DY}^2)$ (eq.(\ref{eq:G-DY-new})).

Being $N_f$-independent, the {\em same} Euclidean or Minkowskian  distribution functions, now fixed
through the large
$N_f$ identification of the Sudakov effective couplings, can then be used to determine the corresponding  
couplings at {\em finite}
$N_f$ in the usual way, requiring  the large $N$ logarithmic terms on the left hand side of
eq.(\ref{eq:scale-viol}) and (\ref{eq:scale-viol-DY}) (with  $G=G_{Mink}$ and $G_{DY}=G_{Mink}^{DY}$ or
$G_{Eucl}^{DY}$)
to be correctly reproduced order by order in
perturbation theory. It is  natural to keep referring to the resulting $A_{{\cal S}}^{Mink}(k^2)$ and
$A_{{\cal S},DY}^{Mink}(k^2)$ couplings (or
$A_{{\cal S}}^{Eucl}(k^2)$ and $A_{{\cal S},DY}^{Eucl}(k^2)$ couplings) 
  as Minkowskian  (resp. Euclidean ) even at {\em finite} $N_f$, where
 identification to a dressed gluon propagator is no longer possible. 
This proposal is equivalent to generalize the basic
equation of the dispersive approach to {\em all orders} in $\alpha_s$ at {\em finite} $N_f$  to the {\em large}
$N$ statement  (in e.g. the DIS case):

\begin{eqnarray}{d\ln F_2(Q^2,N)\over d\ln Q^2}&=4 C_F\int_{0}^{\infty}{d\lambda^2\over \lambda^2} \ddot{{\cal
F}}({\lambda^2\over Q^2},N) A_{{\cal S}}^{Mink}(\lambda^2)\nonumber\\
&+4 C_F \Delta H_{Mink}(Q^2)+{\cal O}({1\over N})\label{eq:dispersive},\end{eqnarray}
where a ``leftover'' $\Delta H_{Mink}(Q^2)$ contribution is still expected at finite $N_f$.

\noindent\underline{IR finite coupling issue}: although
the large $N_f$  coupling $A_{{\cal S},\infty}^{Mink}(k^2)$ of eq.(\ref{eq:A-simple}) is IR finite,
it {\em cannot} be taken as an evidence in favor of the IR finite coupling approach to power corrections at the
{\em perturbative} level, contrary to the statements in \cite{Gru-talk, Gru-talk1}, since it is now clear it should
be identified to a Minkowskian coupling. The corresponding perturbative Euclidean coupling $A_{{\cal
S},\infty}^{Eucl}(k^2)$   is just the one-loop V-scheme coupling (eq.(\ref{eq:one-loop})) 
  which has a Landau pole, and
thus by itself provides no  evidence in favor of the IR finite coupling approach, which relies in an essential way
\cite{DMW,Gru-power} on the IR finitness of the {\em Euclidean} coupling. The {\em assumption} of IR
finitness must therefore be made, as usual \cite{DMW}, at the {\em non-perturbative} level, by postulating the
existence of a non-perturbative modification $\delta A_{{\cal S}}^{Eucl}(k^2)$ of the Euclidean coupling at low
scales. There is nevertheless some (admittedly not yet conclusive) indication for the existence of an IR fixed
point in the perturbative Euclidean coupling at {\em finite} $N_f$. Indeed the three-loop Sudakov effective
coupling beta-function

\begin{equation}{dA_{{\cal S}}^{Eucl}\over d\ln k^2}=-\beta_0 (A_{{\cal S}}^{Eucl})^2-\beta_1 (A_{{\cal
S}}^{Eucl})^3-\beta_2^{Eucl} (A_{{\cal S}}^{Eucl})^4
+...\label{eq:betaeff}\end{equation}
does have a positive zero  $A_{{\cal S},IR}^{Eucl}$ even at low $N_f$ values, due to a large negative
three-loop coefficient
$\beta_2^{Eucl}=-321.0-1508.5 N_f+84.6 N_f^2$ (DIS case). For instance at $N_f=4$ one gets $4\pi A_{{\cal
S},IR}^{Eucl}\simeq 0.6$, which looks marginally perturbative (a similar value is obtained in the Drell-Yan case).
Of course, as in other examples
\cite{G-K}, this fixed point could be easily washed out by 4-loop corrections. However, for $N_f$ close 
to the value $16.5$ where asymptotic freedom is lost, there is  a Banks-Zaks \cite{G-K,Gru-condensate} type of 
fixed point, which is expected to persist within the so-called ``conformal window'', whose lower boundary might
eventually extend down to $N_f$ values as low \cite{Gru-fixing} as $N_f=4$.
Although the standard
Banks-Zaks expansion
 of $A_{{\cal S},IR}^{Eucl}$ in powers of $16.5-N_f$   appears divergent at  $N_f=4$, the
modified expansion suggested in
\cite{Gru-window} yields a reasonably small next-to-leading order correction:

\begin{equation}A_{{\cal S},IR}^{Eucl}\simeq \epsilon (1+5.3 \epsilon+...)\label{eq:new-BZ},\end{equation}
where $\epsilon=(-\beta_0/ \beta_{20}^{Eucl})^{1/2}\simeq 0.06$ at $N_f=4$, with
$\beta_{20}^{Eucl}\equiv\beta_2^{Eucl}\vert N_f=16.5$.
The issue of an IR finite
perturbative Euclidean Sudakov coupling at low values of $N_f$ is thus still open.

\noindent\underline{Universality issues}: apart from its intuitive appeal, the IR finite coupling approach is
potentially more powerful then related approaches \cite{Gru-power,Gardi-Gru,Gardi-review} based on Borel
resummation, since it allows some statement on the universality of power corrections to various processes. Indeed,
I note that at large
$N_f$ there is universality to all orders in $\alpha_s$  between $A_{{\cal S}}^{Eucl}(k^2)$
and
$A_{{\cal S},DY}^{Eucl}(k^2)$, which in the present framework are both prescribed to be equal to the  one-loop
V-scheme coupling in this limit. At finite $N_f$ however it easy to check that  universality in the ultraviolet
region  holds only up to next to leading order in $\alpha_s$, where the DIS and Drell-Yan Euclidean  Sudakov
effective couplings actually coincide
\cite{footnote1} (up to a 
$1/4C_F$ factor) with the ``cusp'' anomalous dimension $A(a_s(k^2))$, but is lost beyond that order.

\noindent On the other hand, an interesting universality property holds in the IR region at finite $N_f$,
{\em assuming} the Sudakov effective couplings reach  non-trivial IR fixed points  at zero momentum: this happens
in particular  within the above mentioned ``conformal window'', but could also take place for lower $N_f$ values
below it, if one assumes \cite{DMW} the couplings are kept  IR finite there through a low energy non-perturbative
modification. Indeed, in the standard resummation formalism  the DIS  and Drell-Yan Sudakov  couplings
differ from the (universal) cusp anomalous dimension only by (non-universal) momentum derivative terms
(eq.(\ref{eq:A-S}) and (\ref{eq:A-S-DY})): they  are  expected  to vanish  at zero momentum, if scale invariance
holds there (i.e. if the
$B$ and $D$ Sudakov anomalous dimensions reach also non-trivial IR fixed points). One thus expects in the standard
scheme equality (up to a 
$1/4C_F$ factor) of  the  DIS and Drell-Yan Sudakov  couplings IR fixed points 
with the cusp IR fixed point value:

\begin{equation}4C_F A_{{\cal S}}^{stan}(0)=4C_F A_{{\cal
S},DY}^{stan}(0)=A(0)\label{eq:fixed-point}.\end{equation} 
Moreover, this property is independent of
the previously discussed ambiguity in the choice of the Sudakov effective couplings and of the variety of Sudakov
resummation procedures. Indeed, eq.(\ref{eq:form-factor}) shows that for a given selection of ``leftover'' constant
terms (contained in the renormalization group invariant
 function
$H(Q^2)$), the corresponding   coupling $A_{{\cal S}}(Q^2)$ differs from the {\em universal}
quantity \cite{footnote2} (for space-like processes)

\begin{equation}A_{{\cal S}}^{all}(Q^2)\equiv -{1\over 4 C_F}{d^2\ln \left({\cal F}_q(Q^2)\right)^2 \over
(d\ln Q^2)^2}\label{eq:AS-all}\end{equation}
only by a total   derivative:

\begin{equation} A_{{\cal S}}(Q^2)=A_{{\cal S}}^{all}(Q^2)+dH/d\ln Q^2\label{eq:form-factor-bis}.\end{equation}
The latter is again expected to vanish at zero momentum if one assumes \cite{footnote3} $H(Q^2)$ also reaches a
non-trivial IR fixed point
$H(0)$. A similar argument applies to 
$A_{{\cal S},DY}(Q^2)$ using eq.(\ref{eq:form-factor-DY}), with $A_{{\cal S}}^{all}(Q^2)$ replaced by its
time-like counterpart $Re[A_{{\cal S}}^{all}(-Q^2)]=-{1\over 4 C_F}{d^2\ln \vert{\cal F}_q(-Q^2)\vert^2 \over
(d\ln Q^2)^2}$. Thus we expect, for {\em any} resummation procedure
(assuming the IR fixed points exist)
 \begin{equation} A_{{\cal S}}(0)= A_{{\cal
S},DY}(0)=A_{{\cal S}}^{all}(0)\label{eq:fixed-point-univ},\end{equation}
 which also reveals \cite{footnote4}, comparing with
eq.(\ref{eq:fixed-point}), that 

\begin{equation}4 C_F A_{{\cal S}}^{all}(0)=A(0)\label{eq:fixed-point-univ1}.\end{equation}
I note that eq.(\ref{eq:fixed-point-univ}) is sufficient to establish IR universality, even if
eq.(\ref{eq:fixed-point}) and (\ref{eq:fixed-point-univ1}) do not hold \cite{footnote5}. These remarks  give some
support to the (approximate) universality  of 
 the corresponding IR power corrections in the IR finite
coupling approach.

In conclusion, the present paper sets the stage for a precise  implementation of the dispersive approach in the
framework of Sudakov resummation. The correct Sudakov distribution functions relevant to the issue of power
corrections have been determined both for DIS and  Drell-Yan. In the Minkowskian representation, they are simply
given by  the corresponding ``characteristic functions'' of
\cite{DMW} (or they (appropriately defined) large $N$ limits).   
The results concerning
power corrections are in agreement with the IR renormalon expectations, but  do not rely on the single gluon
exchange approximation. This procedure should be easily extended to the class of inclusive processes
discussed in
\cite{Sterman-Vogelsang}. It has also been  argued that the assumption that Sudakov effective couplings are  IR
finite implies the universality of the corresponding IR fixed points.

\acknowledgments
I wish to thank M. Beneke, Yu.L. Dokshitzer, J.P. Lansberg, G. Marchesini, G.P. Salam and G. Sterman for early
discussions on the subject of this paper.  I am indebted to S. Friot for the result quoted in 
eq.(\ref{eq:G-DY-Eucl}), and for help in checking eq.(\ref{eq:G-DY-new}).

\bibliography{apssamp}% Produces the bibliography via BibTeX.

\begin{thebibliography}{9}
\bibitem{footnote0} The fact that ${\cal S}$ is a function of only the {\em single} variable $Q^2/N$ (the ``jet
scale'') is the main content of Sudakov resummation in the DIS case.
\bibitem{footnote}   In this sense,  the IR finite coupling ansatz  can  be
seen as a substitute to the  prescription needed \cite{Catani} if a Landau pole is present.
\bibitem{footnote1} This is a non-trivial statement as far as the DIS
coupling $A_{{\cal S}}^{Eucl}(k^2)$ is concerned. In particular, it is not satisfied by the coupling $A_{{\cal
S}}^{stan}(k^2)$ (eq.(\ref{eq:A-S})) resulting from the standard resummation formalism.
\bibitem{footnote4}$A_{{\cal S}}^{all}(0)={d^2\ln \left({\cal F}_q(Q^2)\right)^2\over (d\ln Q^2)^2}\vert( Q^2=0)$
does not necessarily vanish, since ${d \ln \left({\cal F}_q(Q^ 2)\right)^2\over d\ln Q^2}$, although
renormalization group invariant, is not finite in 4 dimensions.
\bibitem{footnote5} The IR behavior of the
cusp anomalous dimension $A(a_s(k^2))$ could be less smooth then that of the ``physical'' Sudakov effective
couplings, as observed in \cite{Gru-talk} at large $N_f$.
\end{thebibliography}

\end{document}